\documentclass[twocolumn,showpacs,preprintnumbers,
amsmath,amssymb]{revtex4}
\usepackage{graphicx}

\begin{document}
\title{Bose-Einstein Condensation Picture of Superconductivity in
$\textrm{YBa}_{2}\textrm{Cu}_{3}\textrm{O}_{7}(91\textrm{K})$ and
$\textrm{YBa}_{2}\textrm{Cu}_{3}\textrm{Se}_{7}(371\,\textrm{K}).$
(Dilute metals).}
\author{V. N. Bogomolov}
\affiliation{A. F. Ioffe Physical \& Technical Institute,\\
Russian Academy of Science,\\
194021 St. Petersburg, Russia} \email{V.Bogomolov@mail.ioffe.ru}
\date{\today}
\begin{abstract}
A metal dilution degree in the compounds
$\textrm{YBa}_{2}\textrm{Cu}_{3}\textrm{O}_{7}$ and
$\textrm{YBa}_{2}\textrm{Cu}_{3}\textrm{Se}_{7}$ is defined as
$z=(r_{\textrm{Me}}/r_{\textrm{O}2-;\textrm{Se}2-})^{3}.$ A
substitution of oxygen by selenium changes \textit{z} by $8$ times
and the Bose-Einstein condensation temperature equals
$T_{\textrm{cSe}} = T_{\textrm{cO}} \cdot8^{2/3} =
364\,\textrm{K}.$ The "active" electron pairs density in
$\textrm{YBa}_{2}\textrm{Cu}_{3}\textrm{O}_{7}$ is about
$1.7\times10^{20} \textrm{cm}^{-3}.$ The electron effective mass
is about $5m_{e}$ and is proportional to the dielectric constant.
\end{abstract}
\pacs{71.30.+h, 74.20.-z, 74.25.Jb}
\maketitle
\bigskip
The superconductivity at $371\,\textrm{K}$ in
$\textrm{YBa}_{2}\textrm{Cu}_{3}\textrm{O}_{7}$ was found
measuring the magnetic properties in \cite{bib1}. However, the
substitution of the oxygen ions $(r_{02-} \sim 0.5 \AA)$ by the
selenium ones $(r_{\textrm{Se}2-} \sim 1.0 \AA)$ made the lattice
unstable and the compound irreproducible.

Densities of the electron pairs, transition temperatures and
effective masses at the "physical" dilution of metals  \\in
$\textrm{Na}_{0.04}\textrm{NH}_{3} \quad (T_{c}\sim 200
\,\textrm{K},\quad m^{*} \sim 5 m_{e})$ \cite{bib2}, \\in
$\textrm{Na}_{0.05}\textrm{WO}_{3}\quad (T_{c}\sim 91
\,\textrm{K},\quad m^{*} \sim 10 m_{e})$ \cite{bib3}, \quad and
\\in
$\textrm{Ag}_{2}(\textrm{Ag}_{3}\textrm{Pb}_{2}\textrm{H}_{2}\textrm{O}_{6})
\; (T_{c}\sim 400 \,\textrm{K},\; m^{*} \sim 7.5 m_{e})$
\cite{bib4} are known and comply with the BEC model \cite{bib5}.

The "active" electron pair density in oxides of the $(1-2-3-7)$
type corresponds to the metal orbital occupancy, which depends
upon the acceptor properties of the oxygen or selenium ions (the
"chemical" dilution) \cite{bib6}. (The "active" electron density
in the Noble Gases condensates determines by interaction of the
atom ground state orbitals \cite{bib6}).

Let us assume that the Me and $\textrm{O}^{2-}$ orbital occupancy
are inversely proportional to the orbital volumes: $z =
(r_{\textrm{Me}}/r_{O2-})^{3}; \quad z = 8.0$ for the pair
$\textrm{Se}^{2-} -- \textrm{O}^{2-}.$ Therefore, the pair density
in YBaCuSe is 8 times larger, than in YBaCuO. For the BEC
mechanism, $T_{\textrm{cSe}} = T_{\textrm{cO}} \cdot8^{2/3} =
 91\cdot4 = 364\,\textrm{K}$ that complies with
the data \cite{bib1}.

About one metal atom falls at one oxygen atom in YBaCuO. The pair
number in the unit cell equals \!$2.5.$ Every electron pair
occupies the volume $69.4 \AA^{3}.$ For the occupancy
$z_{\textrm{Ba--O}} = (2.19/0.5)^{3} = 84$ we have for the
"active" electron pair density is about $1.7\times10^{20}\,
\textrm{cm}^{-3}.$ The transition temperature $T_{c\textrm{O}} =
92\,\textrm{K}$  for the BEC mechanism and the effective mass
$m^{*} \sim 5 m_{e}$ (the dielectric constant $\sim 20$\!\!).

The effective electron  mass magnitudes are  very close  in all of
these four systems \cite{bib1,bib2,bib3,bib4} and comply with the
dielectric constants.

The compounds studied in \cite{bib1} and \cite{bib4} show the
superconductivity of the "gossamer" kind.

A use of hydrogen for blocking of a part of the oxygen valence can
preserve the lattice structure at the electron concentration
increase leading to an increase of $T_{c\textrm{O}}$ \cite{bib7}.

\end{document}